\documentclass[sigplan]{acmart}

\usepackage{booktabs} 

\setcopyright{rightsretained}

\acmConference[FTfJP 2017]{Formal Techniques for Java-like Programs}{June 2017}{Barcelona, Spain} 
\acmYear{2017}
\copyrightyear{2017}

\newcommand{\K}{$\mathbb{K}$}
\newcommand{\KAndroid}{\K Android}
\newcommand{\Kandroid}{\K-Android}

\begin{document}
\pagestyle{empty}

\title{Software Model Checking: \\
A Promising Approach to Verify  Mobile App Security}
\titlenote{This work was funded by EPSRC and received 
advice from Erwin R.\ Catesbeiana (Jr). }
\subtitle{ -- A Position Paper -- }

\author{Irina M\u ariuca As\u avoae}
\affiliation{%
  \institution{INRIA, Paris, France}
}
\email{irina-mariuca.asavoae@inria.fr}

\author{Hoang Nga Nguyen}
\affiliation{%
  \institution{Coventry University, Coventry, UK}
}
\email{hoang.nguyen@coventry.ac.uk}

\author{Markus Roggenbach}
\affiliation{%
  \institution{Swansea University, Swansea, UK}
}
\email{m.roggenbach@swansea.ac.uk}

\author{Siraj Ahmed Shaikh}
\affiliation{%
  \institution{Coventry University, Coventry, UK}
}
\email{siraj.shaikh@coventry.ac.uk}

\renewcommand{\shortauthors}{As\u avoae et al.}

\begin{abstract}
In this position paper we advocate software model checking as a
technique suitable for security analysis of mobile apps. Our
recommendation is based on promising results that we achieved on
analysing app collusion in the context of the Android operating
system. Broadly speaking, app collusion appears when, in performing a
threat, several apps are working together, i.e., they exchange
information which they could not obtain on their own. In this context,
we developed the \Kandroid\ tool, which provides an encoding of the
Android/Smali code semantics within the
\K\ framework. \Kandroid\ allows for software model checking of
Android APK files. Though our experience so far is limited to
collusion, we believe the approach to be applicable to further
security properties as well as other mobile operating systems.
\end{abstract}



\maketitle

\section{Introduction}

\begin{figure}
\centering
\includegraphics[width=0.43\textwidth]{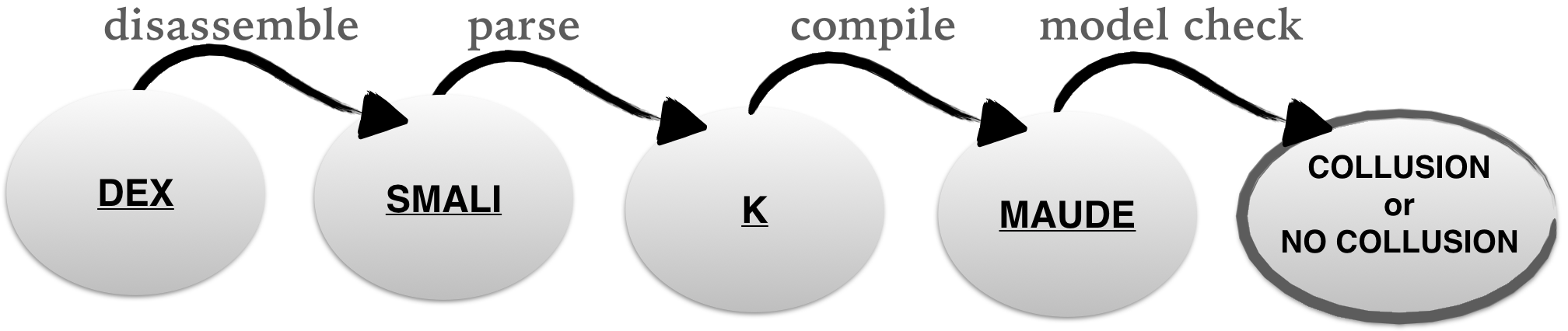}
\caption{Work-flow for model checking with the $\mathbb{K}$ framework.}\label{fig:mc_workflow}
\end{figure}

We advocate as a promising research direction: applying software model
checking to Android apps for formal security analysis. This uses
abstract model checking, which is an abstract interpretation
technique.  Here, we have already achieved a number of explorative
results. These include: defining and experimenting with two executable
semantics on the byte-code level, one concrete and one abstract. Both
of them have been implemented in the \Kandroid\ tool~\cite{ACIDmc16,
  Kandroid}, utilising the
\K\ framework~\cite{rosu-serbanuta-2010-jlap} where Java/JVM semantics
had already been defined~\cite{KJava}.  Our work targets however the
byte-code level and Android operating system (ART/Dalvik); the work-flow of \Kandroid\ is described in Fig.~\ref{fig:mc_workflow}.  Currently
we are pioneering (w.r.t.\ the formal executable semantics for a
virtual machine targeted by Java) a formal proof utilizing a simulation relation that these two
semantics are in a sound relation. 

In the followings we discuss a number of decisions underlying the
suggested approach, give a brief status report on our research, and
conclude by providing some insights that we gained.


{\bf Related work:} Our work is closest to static analysis tools that
detect security properties in Android. For example, the tool
FlowDroid~\cite{flowdroid-pldi14} uses taint analysis to find
connections between source and sink. The app inter-component
communication pattern is subsequently analysed using a composite
constant propagation technique~\cite{octeau16CtPrp}.
We propose a similar approach, namely to track (sensitive) information
flow and to detect app communication, but using model checking that
gives witness traces in case of collusion detection.  From the proof
effort perspective, we mention CompCert~\cite{CompCert} that uses Coq
theorem prover to validate a C compiler.
Also, an up-to-date survey on app collusion in Android can be found in \cite{ACollusionSurvey}.

\section{Decisions}

When setting up our framework for software model checking, we took a
number of decisions that we conceive to be fundamental:

\paragraph{Verify byte-code rather than high level language programs}
When considering the language level, the input language of the virtual
machine appears to be the right level for investigating security
properties. Users download their apps as APKs hence this needs to be
the starting point for our investigation.  Decompiling APKs is a
possibility however not 100\% successful.  A further advantage is that
a language such as Smali, which was designed to run on a Virtual
Machine, is far less complex than a high-level language such as
Java. Finally, Smali programs are independent of compiler
optimisations: verification addressing specific Java constructs might
fail on the byte code level as compiler optimisations might
interfere.

\paragraph{Offer two semantics: a concrete and an abstract one}
We believe it to be essential to work with two different semantics.
Objectives of formulating a concrete semantics include:
\begin{description}
\item[C-O1] To be close to the informal description of the language instructions to ease modelling. For Android these are Smali instructions as specified on the Android Project website~\cite{dex}.
\item[C-O2] To work with actual values as much as possible: this allows to experiment with small example programs in order to validate the given semantics. Note that the \K\ framework allows for executable specifications.
\end{description}
Objectives of formulating an abstract semantics include:
\begin{description}
\item[A-O1] 
To enable effective model checking by selecting suitable abstraction principles. In \Kandroid\ we have chosen:
\begin{itemize}
\item
virtual unrolling: this leads to finite flows~\cite{VIVUtech};
\item
memory abstraction: to reduce the state space~\cite{memoryHard};
\item
constant propagation: this abstracts from concrete values and thus also helps in reducing the state space~\cite{constPropg}.
\end{itemize}
\item[A-O2] To be sound w.r.t.\ the security property under discussion, in our case: collusion. 
\end{description}

\paragraph{Provide a soundness proof}

In order to certify the correctness of the overall approach, a soundness proof is needed. Though the effort required in carrying out such a proof might appear as a high price to pay, the overall setup has a number of advantages:
\begin{itemize}
\item 
The proof is done once; the savings of the abstract semantics in time
and space apply every time model checking is carried out; moreover,
the proof is re-usable as it is structured according to classes of
Smali instructions -- even when changing the property, the abstract
semantics for some of these classes would stay the same.
\item
Working with a single semantics confuses objectives, namely to be true to the informal descriptions (c.f.\ {\bf C-O1} and {\bf C-O2}) and, at the same time to be effective (c.f.\ {\bf A-O1}). This confusion might compromise the overall objective of providing a reliable analysis tool (c.f.\ {\bf A-O2}).
\end{itemize}

\section{Current Status of our work}

\begin{figure}[h]
\centering
\includegraphics[width=0.43\textwidth]{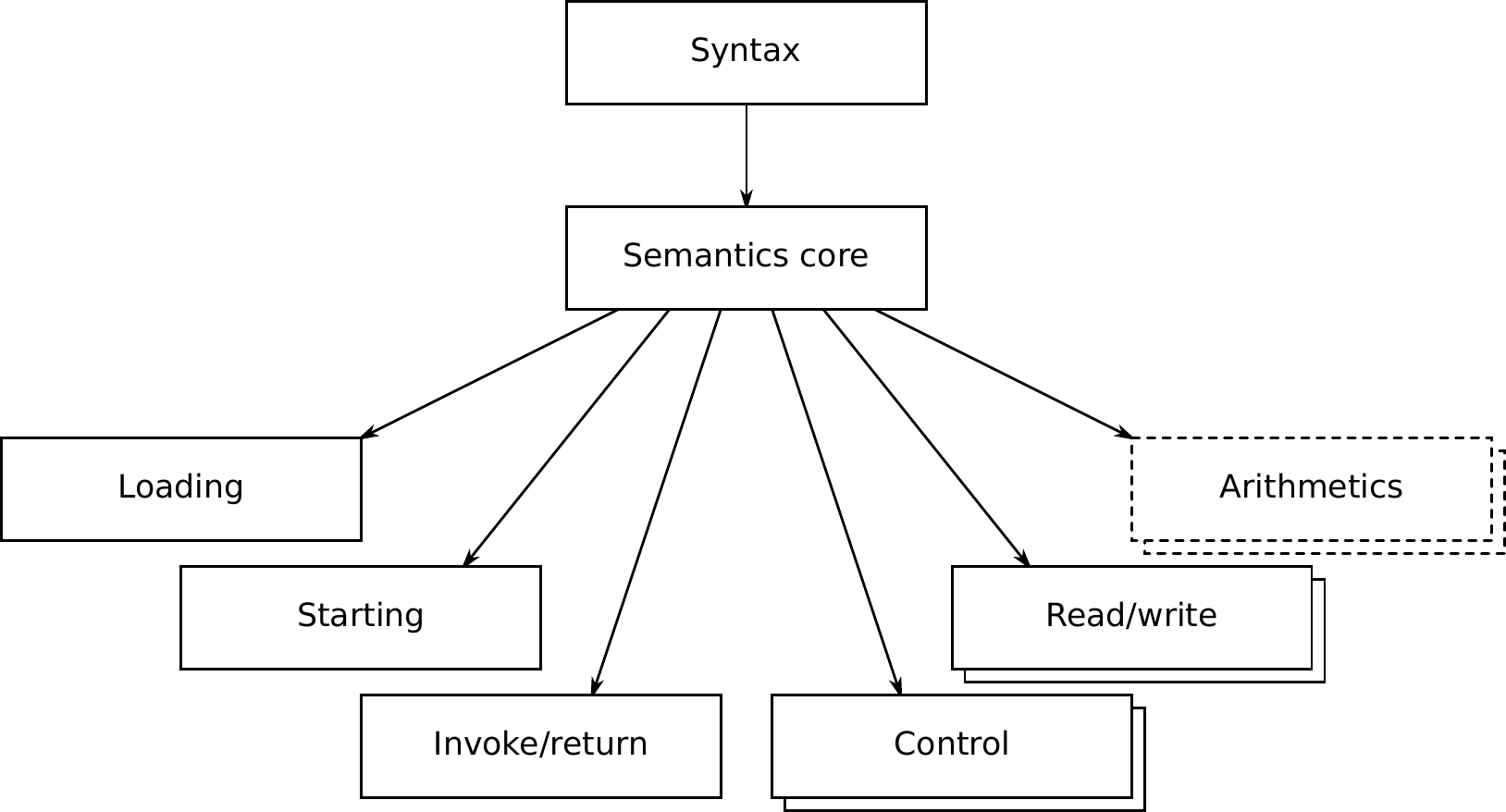}
\caption{Semantic module structure.}
\label{fig:semantics-modules}
\end{figure}

In our tool \Kandroid~\cite{ACIDmc16, Kandroid}, we implement
experimental versions of a concrete and an abstract semantics, which
both cover the whole Smali language--see Figure
\ref{fig:semantics-modules} for the chosen module structure. We have
successfully applied our tool to a number of Android apps to analyse
them for collusion. Here, the counter-example traces provided by the model
checking give good guidance for the code-analysis that
distinguishes between collusion and false positives.

Our correctness proof is "well on its way"--we covered the core constructs, e.g.,
method calls and returns. Although the sheer number of cases to consider (Smali has
about 220 instructions) makes the proof time consuming, we  classified the instructions in about 20  groups that share a similar build. This  modularisation provides the proof with flexibility and reusability characteristics.

\section{First insights}

Concerning the question if it would be possible to directly build a
suitable abstract semantics, our experience suggests that the two step
approach including a proof is a necessity. In our ongoing proof, we
learned that in some cases our originally implemented
semantics went wrong.
Reflecting on the abstraction via a formal simulation relation helped us
to find the correct semantic clauses.

Concerning the applicability of our approach, experiments with our
concrete and abstract semantics indicate that, provided an astute abstraction, software model checking
for security is feasible and might even scale even for demanding properties as collusion.

\section{Conclusion}
Our ongoing work demonstrates that software model checking
is a viable technique for analysing mobile apps for
security. Verification times are below a minute for small examples
consisting of about 5K lines of Smali code.  The concrete semantics
provided as well as the abstraction principles applied can be re-used
to investigate further security properties.  Though \Kandroid\ is
tailored to the Android operating system, the concepts in other mobile
operating systems such as Symbian, MeeGo, iOS, Android, Tizen,
etc.\ appear to be similar enough that it should be possible to apply
software model checking also in their context.  Compared to the
predominant static analysis methods traditionally applied in mobile
security verification, especially the possibility to obtain
counter-example traces makes software model checking a promising
approach.





\bibliographystyle{ACM-Reference-Format}
\bibliography{main}


\begin{thebibliography}{00}


\ifx \showCODEN    \undefined \def \showCODEN     #1{\unskip}     \fi
\ifx \showDOI      \undefined \def \showDOI       #1{{\tt DOI:}\penalty0{#1}\ }
  \fi
\ifx \showISBNx    \undefined \def \showISBNx     #1{\unskip}     \fi
\ifx \showISBNxiii \undefined \def \showISBNxiii  #1{\unskip}     \fi
\ifx \showISSN     \undefined \def \showISSN      #1{\unskip}     \fi
\ifx \showLCCN     \undefined \def \showLCCN      #1{\unskip}     \fi
\ifx \shownote     \undefined \def \shownote      #1{#1}          \fi
\ifx \showarticletitle \undefined \def \showarticletitle #1{#1}   \fi
\ifx \showURL      \undefined \def \showURL       {\relax}        \fi
\providecommand\bibfield[2]{#2}
\providecommand\bibinfo[2]{#2}
\providecommand\natexlab[1]{#1}
\providecommand\showeprint[2][]{arXiv:#2}

\bibitem[\protect\citeauthoryear{{Android Open Source Project}}{{Android Open
  Source Project}}{2016}]%
        {dex}
\bibfield{author}{\bibinfo{person}{{Android Open Source Project}}.}
  \bibinfo{year}{2016}\natexlab{}.
\newblock \bibinfo{title}{{Dalvik Bytecode}}.
\newblock
  \bibinfo{howpublished}{\url{https://source.android.com/devices/tech/dalvik/dalvik-bytecode.html}}.
    (\bibinfo{year}{2016}).
\newblock


\bibitem[\protect\citeauthoryear{Arnold, Manevich, Sagiv, and Shaham}{Arnold
  et~al\mbox{.}}{2006}]%
        {memoryHard}
\bibfield{author}{\bibinfo{person}{Gilad Arnold}, \bibinfo{person}{Roman
  Manevich}, \bibinfo{person}{Mooly Sagiv}, {and} \bibinfo{person}{Ran
  Shaham}.} \bibinfo{year}{2006}\natexlab{}.
\newblock \showarticletitle{Combining Shape Analyses by Intersecting
  Abstractions}. In \bibinfo{booktitle}{{\em {VMCAI} 2006}} {\em
  (\bibinfo{series}{Lecture Notes in Computer Science})},
  Vol.~\bibinfo{volume}{3855}. \bibinfo{publisher}{Springer},
  \bibinfo{pages}{33--48}.
\newblock


\bibitem[\protect\citeauthoryear{Arzt, Rasthofer, Fritz, Bodden, Bartel, Klein,
  Traon, Octeau, and McDaniel}{Arzt et~al\mbox{.}}{2014}]%
        {flowdroid-pldi14}
\bibfield{author}{\bibinfo{person}{Steven Arzt}, \bibinfo{person}{Siegfried
  Rasthofer}, \bibinfo{person}{Christian Fritz}, \bibinfo{person}{Eric Bodden},
  \bibinfo{person}{Alexandre Bartel}, \bibinfo{person}{Jacques Klein},
  \bibinfo{person}{Yves~Le Traon}, \bibinfo{person}{Damien Octeau}, {and}
  \bibinfo{person}{Patrick McDaniel}.} \bibinfo{year}{2014}\natexlab{}.
\newblock \showarticletitle{{FlowDroid:} precise context, flow, field,
  object-sensitive and lifecycle-aware taint analysis for Android apps}. In
  \bibinfo{booktitle}{{\em {ACM} {SIGPLAN} Conference on Programming Language
  Design and Implementation, {PLDI} '14, Edinburgh, United Kingdom - June 09 -
  11, 2014}}. \bibinfo{publisher}{{ACM}}, \bibinfo{pages}{29}.
\newblock


\bibitem[\protect\citeauthoryear{Asavoae, Nguyen, Roggenbach, and
  Shaikh}{Asavoae et~al\mbox{.}}{2016}]%
        {ACIDmc16}
\bibfield{author}{\bibinfo{person}{Irina~Mariuca Asavoae},
  \bibinfo{person}{Hoang~Nga Nguyen}, \bibinfo{person}{Markus Roggenbach},
  {and} \bibinfo{person}{Siraj~Ahmed Shaikh}.} \bibinfo{year}{2016}\natexlab{}.
\newblock \showarticletitle{Utilising {K} Semantics for Collusion Detection in
  Android Applications}. In \bibinfo{booktitle}{{\em FMICS-AVoCS 2016}} {\em
  (\bibinfo{series}{Lecture Notes in Computer Science})},
  Vol.~\bibinfo{volume}{9933}. \bibinfo{publisher}{Springer},
  \bibinfo{pages}{142--149}.
\newblock


\bibitem[\protect\citeauthoryear{Bhandari, Jaballah, Jain, Laxmi, Zemmari,
  Gaur, and Conti}{Bhandari et~al\mbox{.}}{2016}]%
        {ACollusionSurvey}
\bibfield{author}{\bibinfo{person}{Shweta Bhandari}, \bibinfo{person}{Wafa~Ben
  Jaballah}, \bibinfo{person}{Vineeta Jain}, \bibinfo{person}{Vijay Laxmi},
  \bibinfo{person}{Akka Zemmari}, \bibinfo{person}{Manoj~Singh Gaur}, {and}
  \bibinfo{person}{Mauro Conti}.} \bibinfo{year}{2016}\natexlab{}.
\newblock \showarticletitle{Android App Collusion Threat and Mitigation
  Techniques}.
\newblock \bibinfo{journal}{{\em CoRR\/}}  \bibinfo{volume}{abs/1611.10076}
  (\bibinfo{year}{2016}).
\newblock
\showURL{%
\url{http://arxiv.org/abs/1611.10076}}


\bibitem[\protect\citeauthoryear{Bogdanas and Rosu}{Bogdanas and Rosu}{2015}]%
        {KJava}
\bibfield{author}{\bibinfo{person}{Denis Bogdanas} {and}
  \bibinfo{person}{Grigore Rosu}.} \bibinfo{year}{2015}\natexlab{}.
\newblock \showarticletitle{K-Java: {A} Complete Semantics of Java}. In
  \bibinfo{booktitle}{{\em {POPL} 2015}}. \bibinfo{publisher}{{ACM}},
  \bibinfo{pages}{445--456}.
\newblock


\bibitem[\protect\citeauthoryear{{Kandroid ACID Team}}{{Kandroid ACID
  Team}}{2017}]%
        {Kandroid}
\bibfield{author}{\bibinfo{person}{{Kandroid ACID Team}}.}
  \bibinfo{year}{2017}\natexlab{}.
\newblock \bibinfo{title}{{Kandroid Tool}}.
\newblock
  \bibinfo{howpublished}{\url{http://www.cs.swan.ac.uk/~csmarkus/ProcessesAndData/androidsmali-semantics-k}}.
    (\bibinfo{year}{2017}).
\newblock


\bibitem[\protect\citeauthoryear{Kinder, Zuleger, and Veith}{Kinder
  et~al\mbox{.}}{2009}]%
        {constPropg}
\bibfield{author}{\bibinfo{person}{Johannes Kinder}, \bibinfo{person}{Florian
  Zuleger}, {and} \bibinfo{person}{Helmut Veith}.}
  \bibinfo{year}{2009}\natexlab{}.
\newblock \showarticletitle{An Abstract Interpretation-Based Framework for
  Control Flow Reconstruction from Binaries}. In \bibinfo{booktitle}{{\em
  {VMCAI} 2009}} {\em (\bibinfo{series}{Lecture Notes in Computer Science})},
  Vol.~\bibinfo{volume}{5403}. \bibinfo{publisher}{Springer},
  \bibinfo{pages}{214--228}.
\newblock


\bibitem[\protect\citeauthoryear{Leroy}{Leroy}{2009}]%
        {CompCert}
\bibfield{author}{\bibinfo{person}{Xavier Leroy}.}
  \bibinfo{year}{2009}\natexlab{}.
\newblock \showarticletitle{Formal verification of a realistic compiler}.
\newblock \bibinfo{journal}{{\em Commun. {ACM}\/}} \bibinfo{volume}{52},
  \bibinfo{number}{7} (\bibinfo{year}{2009}), \bibinfo{pages}{107--115}.
\newblock


\bibitem[\protect\citeauthoryear{Martin, Alt, Wilhelm, and Ferdinand}{Martin
  et~al\mbox{.}}{1998}]%
        {VIVUtech}
\bibfield{author}{\bibinfo{person}{Florian Martin}, \bibinfo{person}{Martin
  Alt}, \bibinfo{person}{Reinhard Wilhelm}, {and} \bibinfo{person}{Christian
  Ferdinand}.} \bibinfo{year}{1998}\natexlab{}.
\newblock \showarticletitle{Analysis of Loops}. In \bibinfo{booktitle}{{\em
  Compiler Construction, 7th International Conference in ETAPS'98}} {\em
  (\bibinfo{series}{Lecture Notes in Computer Science})},
  Vol.~\bibinfo{volume}{1383}. \bibinfo{publisher}{Springer},
  \bibinfo{pages}{80--94}.
\newblock


\bibitem[\protect\citeauthoryear{Octeau, Luchaup, Jha, and McDaniel}{Octeau
  et~al\mbox{.}}{2016}]%
        {octeau16CtPrp}
\bibfield{author}{\bibinfo{person}{Damien Octeau}, \bibinfo{person}{Daniel
  Luchaup}, \bibinfo{person}{Somesh Jha}, {and} \bibinfo{person}{Patrick~D.
  McDaniel}.} \bibinfo{year}{2016}\natexlab{}.
\newblock \showarticletitle{Composite Constant Propagation and its Application
  to Android Program Analysis}.
\newblock \bibinfo{journal}{{\em {IEEE} Trans. Software Eng.\/}}
  \bibinfo{volume}{42}, \bibinfo{number}{11} (\bibinfo{year}{2016}),
  \bibinfo{pages}{999--1014}.
\newblock


\bibitem[\protect\citeauthoryear{Ro{\c s}u and {\c S}erb{\u a}nu{\c t}{\u
  a}}{Ro{\c s}u and {\c S}erb{\u a}nu{\c t}{\u a}}{2010}]%
        {rosu-serbanuta-2010-jlap}
\bibfield{author}{\bibinfo{person}{Grigore Ro{\c s}u} {and}
  \bibinfo{person}{Traian~Florin {\c S}erb{\u a}nu{\c t}{\u a}}.}
  \bibinfo{year}{2010}\natexlab{}.
\newblock \showarticletitle{An Overview of the {K} Semantic Framework}.
\newblock \bibinfo{journal}{{\em Journal of Logic and Algebraic Programming\/}}
  \bibinfo{volume}{79}, \bibinfo{number}{6} (\bibinfo{year}{2010}),
  \bibinfo{pages}{397--434}.
\newblock


\end{thebibliography}

\end{document}